\documentclass[prl,aps,twocolumn,showpacs,superscriptaddress]{revtex4}
\usepackage{amsbsy}
\usepackage{bm}
\usepackage{graphicx}

\newcommand{\nc}{\newcommand}
\nc{\noi}{\noindent}     
\nc{\eq}[1]{\mbox{Eq.~(\ref{#1})}}
\nc{\ba}{\begin{array}}
\nc{\ea}{\end{array}}
\nc{\bea}{\begin{eqnarray}}
\nc{\eea}{\end{eqnarray}}
\nc{\fig}[1]{\mbox{Fig.~\ref{#1}}}

\begin{document}

\title{Spin Decoherence in Superconducting Atom Chips}

\author{Per Kristian Rekdal}\email{per.rekdal@uni-graz.at}
\affiliation{Institut f\"ur Physik, Karl-Franzens-Universit\"at Graz, Universit\"atsplatz 5, A-8010 Graz, Austria}
\author{Bo-Sture K. Skagerstam}\email{boskag@phys.ntnu.no}
\affiliation{Complex Systems and Soft Materials Research Group, Department of Physics, 
             The Norwegian University of Science and Technology, N-7491 Trondheim, Norway}
\author{Ulrich Hohenester}
\affiliation{Institut f\"ur Physik, Karl-Franzens-Universit\"at Graz, Universit\"atsplatz 5, A-8010 Graz, Austria}
\author{Asier Eiguren}
\affiliation{Institut f\"ur Physik, Karl-Franzens-Universit\"at Graz, Universit\"atsplatz 5, A-8010 Graz, Austria}

%\date{\today}

\begin{abstract}

     Using a consistent quantum-mechanical treatment for the electromagnetic radiation,
     we theoretically investigate the magnetic spin-flip scatterings of a neutral two-level atom trapped
     in the vicinity of a superconducting body. We derive a simple scaling law for 
     the corresponding spin-flip lifetime for such an atom trapped near a superconducting thick slab.
     For temperatures below the superconducting transition temperature $T_c$, the lifetime
     is found to be enhanced by several orders of magnitude in comparison to the case of a normal conducting slab.
     At zero temperature the spin-flip lifetime is given by the unbounded free-space value.

\end{abstract}


\maketitle

     Coherent manipulation of matter waves is one of the ultimate goals of atom optics.
     Trapping and manipulating cold neutral atoms in microtraps near surfaces of atomic
     chips is a promising approach towards full control of matter waves on small scales \cite{folman_02}.
     The subject of atom optics is making rapid progress, driven both by the fundamental
     interest in quantum systems and by the prospect of new devices based on quantum
     manipulations of neutral atoms.

     With lithographic or other surface-patterning processes complex atom chips can be built
     which combine many traps, waveguides, and other elements, in order to realize 
     controllable composite quantum systems \cite{zoller_00} as needed, e.g., for the implementation of quantum
     information devices \cite{divin_00}.
     Such microstructured surfaces have been highly successful and form the basis of a growing number of experiments
     \cite{hommelhoff_05}. 
     However, due to the proximity of the cold atom cloud to the macroscopic substrate additional
     decoherence channels are introduced which limit the performance of such atom chips.
     Most importantly, Johnson noise currents in the material cause electromagnetic field fluctuations 
     and hence threaten to decohere the quantum state of the atoms. 
     This effect arises because the finite temperature and resistivity of the surface material are always accompanied
     by field fluctuations, as a consequence of the fluctuation-dissipation theorem. 
     Several experimental \cite{hinds_03,vuletic_04,harber_03} as well as theoretical \cite{henkel_99,dung_00,rekdal_04,scheel_05} 
     studies have recently shown that rf spin-flip transitions are the main source of decoherence for atoms 
     situated close to metallic or dielectric bodies.
     Upon making spin-flip transitions, the atoms become more weakly trapped or even lost from the microtrap.

     In Ref. \cite{rekdal_04} it was shown that to reduce the spin decoherence of atoms outside a metal in the normal state, 
     one should avoid materials whose skin depth at the spin-flip transition frequency is comparable with
     the atom-surface distance.
     For typical values of these parameters used in experiments, however,
     this worst-case scenario occurs \cite{hinds_03,vuletic_04,harber_03}.
     To overcome this deficiency, it was
     envisioned \cite{scheel_05} that superconductors might be beneficial in this respect because of their efficient 
     screening properties, although this conclusion was not backed by a proper theoretical analysis. It is the purpose of this
     letter to present a consistent theoretical description of atomic spin-flip transitions in the vicinity of superconducting
     bodies, using a proper quantum-mechanical treatment for the electromagnetic radiation, and to reexamine Johnson-noise 
     induced decoherence for superconductors. We find that below the superconducting transition temperature $T_c$ the spin-flip
     lifetime becomes boosted by several orders of magnitude, a remarkable finding which is attributed to: (1) the opening of the 
     superconducting gap and the resulting inability to deposit energy into the superconductor, (2) the highly efficient 
     screening properties of superconductors, and (3) the small active volume within which current fluctuations can contribute 
     to field fluctuations. Our results thus suggest that current-noise induced decoherence in atomic chips can be completely 
     diminished by using superconductors instead of normal metals.

     We begin by considering an atom in an initial state $|i \rangle$ and trapped at position ${\bf r}_A$
     in vacuum, near a dielectric body. The rate of spontaneous and thermally stimulated magnetic spin-flip
     transition into a final state $|f\rangle$ has been derived in Ref. \cite{rekdal_04},

\bea \label{gamma_B_generel}
     \lefteqn{\Gamma^B = \, \mu_0 \,  \frac{2 \, (\mu_B g_S)^2}{\hbar} \; \sum_{j,k=1}^3 ~  \langle f|\hat{S}_j|i \rangle \langle i|\hat{S}_k|f \rangle }
     \nonumber 
     \\ 
     && \; \times \;
     \mbox{Im}  \, [  \; \nabla \times \nabla \times
     \bm{G}({\bf r}_A  ,  {\bf r}_A  ,  \omega ) \; ]_{jk}  \;  ( \overline{n}_{\text{th}} + 1 ) \, .
\eea

     \noi
     Here $\mu_B$ is the Bohr magneton, $g_s \approx 2$ is the electron spin $g$ factor,
     $\langle f|\hat{S}_j|i\rangle$ is the matrix element of the electron spin operator
     corresponding to the transition $|i\rangle \rightarrow |f\rangle$, and $\bm{G}({\bf r}_A  ,  {\bf r}_A  ,  \omega )$
     is the dyadic Green tensor of Maxwell's theory. 
     \eq{gamma_B_generel} follows from a consistent quantum-mechanical treatment of electromagnetic 
     radiation in the presence of absorbing bodies \cite{dung_00,henry_96}. 
     Thermal excitations of the electromagnetic field modes are accounted for by the factor 
     $( \overline{n}_{\text{th}} + 1 )$, where $\overline{n}_{\text{th}} = 1 / ( e^{\hbar \omega/k_{\text{B}} T}- 1 )$
     is the mean number of thermal photons per mode at frequency $\omega$ of the spin-flip transition. The dyadic 
     Green tensor is the unique solution to the Helmholtz equation

\bea   \label{G_Helm}
   \nabla\times\nabla\times \bm{G}({\bf r},{\bf r}',\omega) - k^2
   \epsilon({\bf r},\omega) \bm{G}({\bf r},{\bf r}',\omega) = \delta( {\bf r} - {\bf r}' ) \bm{1} \, ,\nonumber\\
\eea

     \noi
     with appropriate boundary conditions. Here $k=\omega/c$ is the wavenumber in vacuum, $c$ is the speed of light 
     and $\bm{1}$ the unit dyad. This quantity contains all 
     relevant information about the geometry of the material and, through the electric permittivity
     $\epsilon({\bf r},\omega)$, about its dielectric properties.

     The current density in superconducting media is commonly described by the Mattis Bardeen theory \cite{mattis_58}.
     To simplify the physical picture, let us limit the discussion to low but non-zero frequencies
     $0 < \omega \ll \omega_g \equiv 2 \Delta(0)/\hbar$, where $\omega$ is the angular frequency and
     $\Delta(0)$ is the energy gap of the superconductor at zero temperature.
     In this limit, the current density is well described by means of a two-fluid model \cite{gorter_34,london_34_40}. 
     At finite temperature $T$, the current density consists of two types of electrons, superconducting electrons and
     normal conducting electrons. The total current density is equal to the sum of a superconducting
     current density and a normal conducting current density, i.e.
     ${\bf J}({\bf r}, t) = {\bf J}_s({\bf r}, t) + {\bf J}_n({\bf r}, t)$.
     Let us furthermore assume that the superconducting as well as the normal conducting part of the current density
     responds linearly and locally to the electric field \cite{remark_local},
     in which case the current densities are given by the London equation and Ohm's law,
     respectively,

\bea  \label{J_London}
  \frac{ \partial {\bf J}_s({\bf r},t)}{\partial t} =  \frac{{\bf E}({\bf r},t)}{\mu_0 \, \lambda_L^2(T) } 
  ~~~  ,  ~~~ 
  {\bf J}_n({\bf r},t)  = \sigma_n(T) \, {\bf E}({\bf r},t) \, .
\eea

   \noi
   The London penetration length and the normal conductivity are given by, % respectively, 

\bea  \label{lambda_sigma_def}
  \lambda_L^2(T)  &=&  \frac{m}{\mu_0 \, n_s(T) \, e^2} ~~~ , ~~~  \sigma_n(T)  =  \frac{n_n(T)}{n_0} \, \sigma ~ .
\eea

     \noi
     Here $\sigma$ is the electrical conductivity of the metal in the normal state, 
     $m$ is the electron mass, $e$ is the electron charge, and $n_s(T)$ and $n_n(T)$ are the electron densities
     in the superconducting and normal state, respectively, at a given temperature $T$. Following 
     London \cite{london_34_40}, we assume that the total density is constant and given by
     $n_0 = n_s(T) + n_n(T)$, where $n_s(T)=n_0$ for $T=0$ and $n_n(T)=n_0$ for $T > T_c$. 
     For a London superconductor with the assumptions as mentioned above, the dielectric function 
     $\epsilon(\omega)$ in the low-frequency regime reads

\bea \label{eps_j}
  \epsilon(\omega) = 1 - \frac{1}{k^2 \lambda_L^2(T)} + i \, \frac{2}{k^2 \delta^2(T)} \, ,
\eea

    \noi
    where $\delta(T) \equiv \sqrt{2/\omega \mu_0 \, \sigma_n(T)}$ is the skin depth associated with the
    normal conducting electrons. The optical conductivity corresponding to \eq{eps_j} is
    $\sigma(T) = 2/\omega \mu_0 \delta^2(T) + i/\omega \mu_0 \lambda_L^2(T)$.

\begin{figure}%[ht]

\begin{picture}(0,0)(78,286)  

\centerline{\includegraphics[width=10.0cm]{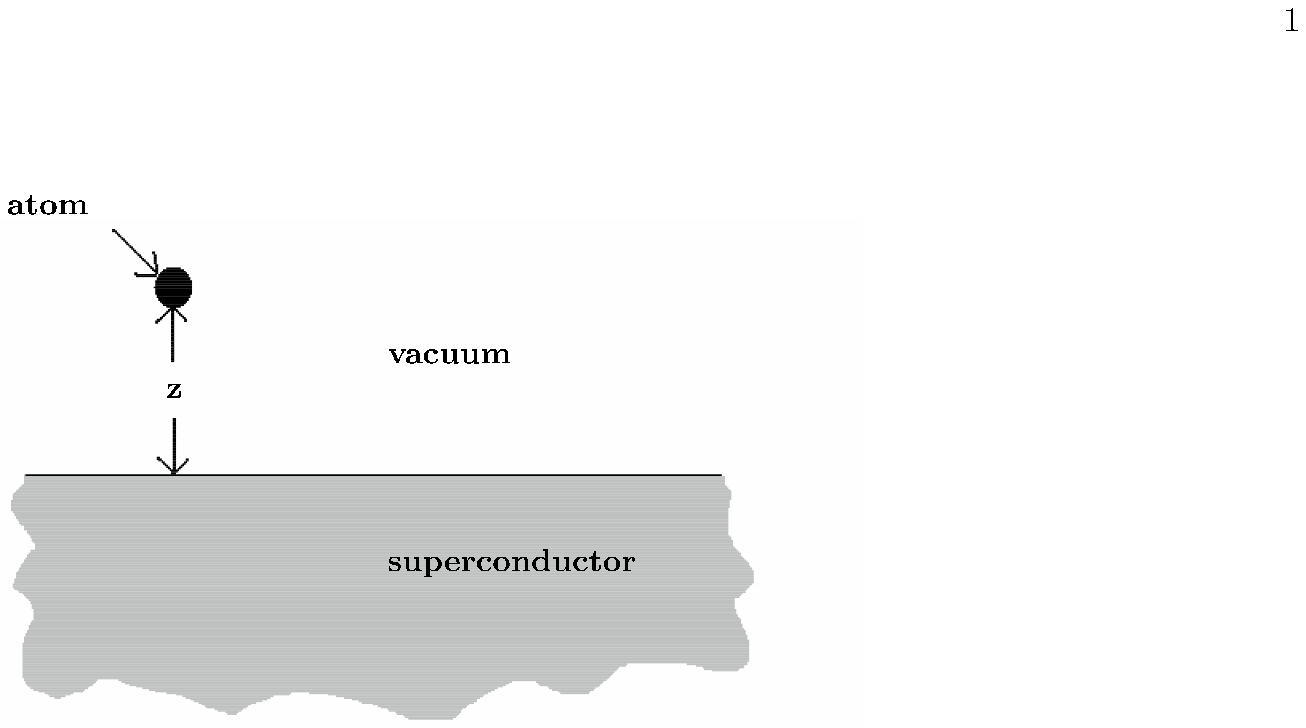}}

\end{picture}

\vspace{4cm}

\caption{Schematic picture of the setup considered in our calculations. An atom inside a magnetic microtrap is 
         located in vacuum at a distance $z$ away from a thick superconducting slab, i.e. a semi-infinite plane.
         Upon making a spin-flip transition, the atom becomes more weakly trapped and is eventually lost.}
\label{geo_slab_fig}
\end{figure}

    In the following we apply our model to the geometry shown in \fig{geo_slab_fig}, where an atom is located in vacuum
    at a distance $z$ away from a superconducting slab. We consider, in correspondence to recent experiments 
    \cite{vuletic_04,harber_03,hinds_03}, $^{87}$Rb atoms that are initially
    pumped into the $|5 S_{1/2},F=2,m_F=2\rangle \equiv |2,2\rangle$ state.
    Fluctuations of the magnetic field may then cause the atoms to evolve into hyperfine sublevels
    with lower $m_F$. Upon making a spin-flip transition to the $m_F=1$ state, the atoms are more weakly trapped
    and are largely lost from the region of observation, causing the measured atom number to decay
    with rate $\Gamma^B_{21}$ associated with the rate-limiting transition $|2,2 \rangle \rightarrow |2,1 \rangle$.
    The transition rate $\Gamma^B_{21} =  ( \Gamma^0_{21} +  \Gamma^{\, \rm{slab}}_{21} ) \, ( \overline{n}_{th} + 1 )$ 
    can be decomposed into a free part and a part purely due to the presence of the slab.
    The free-space spin-flip rate at zero temperature is 
    $\Gamma^{\; 0}_{ 12 } =  \, \mu_0  \, \frac{ ( \mu_B g_S )^2 }{24 \pi  \, \hbar} \, k^3$ \cite{rekdal_04}.
    The slab-contribution can be obtained by matching the electromagnetic fields at the vacuum-superconductor interface. 
    With the same spin-orientation as in Ref.\cite{scheel_05}, i.e. 
    $|\langle f|\hat{S}_y|i\rangle|^2 = |\langle f|\hat{S}_z|i\rangle|^2$ and $\langle f|\hat{S}_x|i\rangle=0$,
    the spin-flip rate is $\Gamma^{\, \rm{slab}}_{21} = \Gamma^0_{21}  ( \,  
    \widetilde{I}_{\|} + \widetilde{I}_{\perp} \, )$, with the atom-spin orientation dependent integrals

\bea  \label{I_paral}
  \widetilde{I}_{\|}    &=&
                \frac{3}{8}   
                {\rm Re} 
                \bigg ( 
        \int_{0}^{\infty} d q  \,  \frac{q}{\widetilde{\eta}_0}  \, e^{ i \, 2 \widetilde{\eta}_0 \, k z } \, 
        [  \, r_p(q)   -  \widetilde{\eta}_0^2  \, r_s(q)  \, ]  \bigg ) \, , ~~~ 
\\ \nonumber
\\
   \label{I_perp}
   \widetilde{I}_{\perp}  &=& 
                \frac{3}{4}  
                {\rm Re} 
                \bigg ( \,
                %s
                \int_{0}^{\infty}  d q  \, \frac{q^3}{\widetilde{\eta}_0} \, e^{ i \, 2 \widetilde{\eta}_0 \, k z } \, 
                r_s(q) \, \bigg ) ,
\eea

   \noi
   and the electromagnetic field polarization dependent Fresnel coefficients

\bea \label{r_s_og_r_p}
  r_s(q)   = \frac{\widetilde{\eta}_0 - \widetilde{\eta}(\omega)}{\widetilde{\eta}_0 + \widetilde{\eta}(\omega)} ~~ , ~~
  r_p(q)   = 
                   \frac{\epsilon(\omega) \, \widetilde{\eta}_0 - \widetilde{\eta}(\omega)}
                        {\epsilon(\omega) \, \widetilde{\eta}_0 + \widetilde{\eta}(\omega)} \, .
\eea

   \noi 
   Here we have ${\widetilde{\eta}}(\omega)  =  \sqrt{\epsilon(\omega) - q^2}$ and  ${\widetilde{\eta}}_0 = \sqrt{1 - q^2}$.
   In particular, above the transition temperature $T_c$ the dielectric function in \eq{eps_j} reduces to the well known
   Drude form.
   Due to the efficient screening properties of superconductors, in most cases of interest the inequality 
   $\lambda_L(T) \ll  \delta(T)$ holds. 
   Assuming furthermore the near-field case $\lambda_L(T) \ll z \ll \lambda$, where $\lambda = 2\pi/k$ is the wavelength
   associated to the spin-flip transition, which holds true in practically all cases of interest, we can compute 
   the integrals in Eqs. (\ref{I_paral}--\ref{r_s_og_r_p}) analytically to finally obtain

\bea   \label{lifetime}
  \Gamma^B_{21}  &\approx&  \Gamma^0_{21}  ( \overline{n}_{th} + 1 )  
                      \left [  1   +    2 ( \frac{3}{4} )^3   \frac{1}{k^3\delta(T)^2} 
                                           \frac{\lambda_L^3(T)}{z^4}  \right ] \, . ~~~
\eea

    \noi
    For a superconductor at $T=0$, in which case there are no normal conducting electrons, 
    it is seen from \eq{lifetime} that the lifetime is given by the unbounded free-space lifetime $\tau_0 \equiv 1/\Gamma^0_{21}$.

    \eq{lifetime} is the central result of our paper. To inquire into its details, we compute the spin-flip
    rate for the superconductor niobium (Nb) and for a typical atomic transition frequency $\nu \equiv \omega/2\pi = 560$ kHz \cite{hinds_03}.
    We keep the atom-surface distance fixed at $z=50\,\mu$m, and use the Gorter-Casimir \cite{gorter_34} temperature dependence

\begin{figure}

\centerline{\includegraphics[width=1.0\columnwidth]{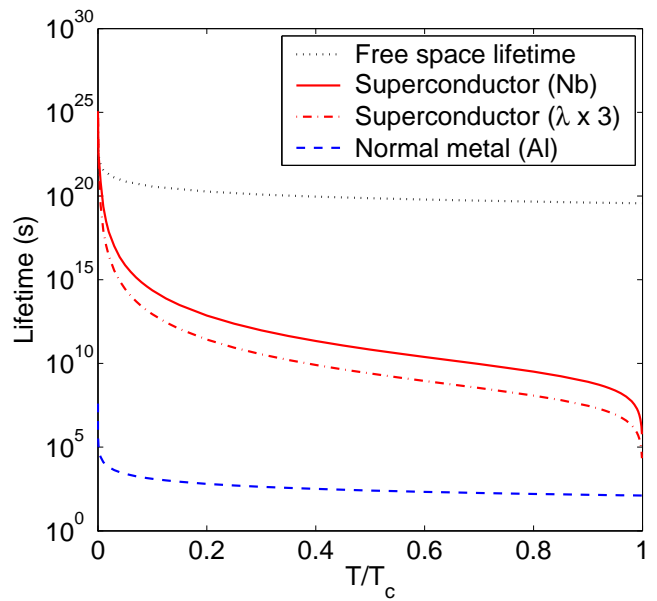}}

\caption{Spin-flip lifetime of a trapped atom near a superconducting slab $\tau_s$ (red solid line) as a function of temperature $T$.
         The atom-surface distance is fixed at $z=50\, \mu$m, and the frequency of the atomic transition is $\nu = 560 \,$kHz.
         The other parameters are $\lambda_L(0) = 35 \,$nm \cite{pronin_98}, 
         $\sigma \approx 2 \cdot 10^9\, (\Omega \mbox{m})^{-1}$ \cite{casalbuoni_05}, and $T_C=8.31$ K \cite{pronin_98}, 
         corresponding to superconducting Nb. 
         The numerical value of $\tau_s$ is computed using the temperature dependence as given by
         Eq. (\ref {n_s_temp_dep}). 
         As a reference, we have also plotted the lifetime $\tau_n$ (blue dashed line) for an atom outside a normal conducting slab 
         with $\delta = 110 \, \mu$m, corresponding to Al.
         The red dashed-dotted line is the lifetime for the same parameters as mentioned above but $\lambda_L(0) = 3 \times 35 \,$nm,
         i.e., where we have taken into account the fact that the London length is modified due to non-local effects.
         The dotted line corresponds to the lifetime $\tau_0/( \overline{n}_{\text{th}} + 1 )$ for a perfect normal conductor.
         The unbounded free-space lifetime at zero temperature is $\tau_0  \approx 10^{25}$ s.}
\label{tau_t_super_og_normal}
\end{figure}

\bea  \label {n_s_temp_dep}
  \frac{n_s(T)}{n_0} = 1 - \frac{n_n(T)}{n_0} = 1 - \left ( \frac{T}{T_C} \right )^4  \, ,
\eea

    \noi for the superconducting electron density. Figure \ref{tau_t_super_og_normal} shows the spin-flip lifetime 
    $\tau_s \equiv 1/\Gamma_{21}^B$ of the atom as a function of temperature: over a wide temperature range $\tau_s$ remains 
    as large as $10^{10}$ seconds. In comparison
    to the normal-metal lifetime $\tau_n$, which is obtained for aluminum with its quite small skin depth $\delta=110 \, \mu$m  and 
    using the results of Refs.~\cite{rekdal_04,scheel_05}, we observe that the lifetime becomes boosted 
    by almost ten orders of magnitude in the superconducting state. In particular, for $T=0$ the ratio between $\tau_s$ and
    $\tau_n$ is even $10^{17}$.
    From the scaling behavior (\ref{lifetime}) we thus observe that decoherence induced by current-fluctuations in the 
    superconducting state remains completely negligible even for small atom-surface distances around 1 $\mu$m, in strong contrast 
    to the normal state where such decoherence would limit the performance of atomic chips.

    The scaling behavior of the spin-flip rate (\ref{lifetime}) can be understood qualitatively on the basis of 
    Eq. (\ref{gamma_B_generel}). The fluctuation-dissipation theorem \cite{dung_00,henry_96} relates the 
    imaginary part of the Green tensor and $\epsilon(\omega)$ by ${\rm Im} \, {\bf G} = {\bf G} \, {\rm Im} [\epsilon(\omega) ] \, {\bf G}^*$, 
    assuming a suitable real-space convolution, and allows to bring the scattering rate Eq. (\ref{gamma_B_generel})
    to a form reminiscent of Fermi's golden rule. The magnetic dipole of the atom at ${\bf r}_A$ couples to a current fluctuation
    at point ${\bf r}$ in the superconductor through $\bm{G}({\bf r}_A,{\bf r},\omega )$.
    The propagation of the current fluctuation is described by the dielectric function $\epsilon(\omega)$, and finally a
    back-action on the atomic dipole occurs via $\bm{G}({\bf r},{\bf r}_A,\omega )$.
    For the near-field coupling
    under consideration, $z\ll\lambda$, the dominant contribution of the Green tensor is $|{\bf G}| \sim 1/z^2$, thus resulting in the
    overall $z^{-4}$ dependence of the spin-flip rate (\ref{lifetime}). The imaginary part ${\rm Im} \, [ \epsilon(\omega) ] \sim 1/ \delta^2$
    of the dielectric function (\ref{eps_j}) accounts for the loss of electromagnetic energy to the superconductor, and is only governed by
    electrons in the normal state, whereas electrons in the superconducting state cannot absorb energy because of the superconducting gap.
    Finally, the term $\lambda^3$ is due to the dielectric screening $1/\epsilon(\omega) \sim \lambda^2$ of the charge fluctuation seen
    by the atom, and an additional $\lambda$ contribution associated to the active volume of current fluctuations which contribute to the
    magnetic field fluctuations at the position of the atom. Fluctuations deeper inside the superconductor are completely screened out.
    In comparison to the corresponding scaling $\Gamma^B \sim \delta/z^4$ for a normal metal \cite{scheel_05}, which can be qualitatively
    understood by a similar reasoning, the drastic lifetime enhancement in the superconducting state is thus due to the combined effects of
    the opening of the superconducting gap, the highly efficient screening, and the small active volume.

     Let us finally briefly comment on the validity of our simplified approach, and how our results would be
     modified if using a more refined theory for the description of the superconductor. Our theoretical approach is 
     valid in the same parameter regime as London's theory, that is $\lambda(T)\gg \xi(T)$. 
     It is well known that nonlocal effects modify the London length in Nb from 
     $\lambda_L(0)\approx 35$ nm to $\lambda(0) \approx 90$ nm \cite{miller_59}, and the coherence length $\xi(T)$, 
     according to Pippard's theory \cite{pippard_53}, from the BCS value $\xi_0$ to $1/\xi(T) = 1/\xi_0 + 1/\alpha \ell(T)$, where $\alpha$
     is of the order one and $\ell(T)$ is the mean free path. For Nb, $\xi_0=39\,$nm and 
     $\ell(T \le 9 K ) \cong 9\,$nm \cite{pronin_98}, 
     and the London condition $\lambda(T) \gg \xi(T)$ is thus satisfied. Furthermore, at the atomic transition
     frequency the conductivity is $\sigma \approx 2 \cdot 10^9\, (\Omega \mbox{m})^{-1}$ \cite{casalbuoni_05} and the corresponding skin
     depth is $\delta=\sqrt{2/\omega \mu_0 \sigma} \approx 15 \, \mu{\rm m} \leq \delta(T)$, such that Ohm's law is also valid since
     $\delta(T) \gg \ell(T)$ \cite{reuter_48}. It is important to realize that other possible modifications of the parameters used
     in our calculations will by no means change our findings, which only rely on the generic superconductor properties
     of the efficient screening and the opening of the energy gap, and that our conclusions will also prevail for other superconductor materials.

     We also mention that for both a superconductor at $T=0$ and a perfect normal conductor, i.e. $\delta=0$, the lifetime is 
     given by the unbounded free-space lifetime $\tau_0$.
     In passing, we notice that for an electric dipole transition and for a perfect normal conductor,  as e.g discussed in
     Refs.\cite{knight_73}, the correction to the vacuum rate is in general opposite in sign as compared to 
     that of a magnetic dipole transition.
     Elsewhere decay processes in the vicinity of a thin superconducting film will discussed in detail \cite{rekdal_07}.

%     For an electric dipole transition with the atomic dipole orientation such that $d_y^2 = d_z^2$ and $d_x=0$, 
%     the corrections to the vacuum rate is opposite in sign as compared to the (see e.g. \cite{knight_73}).

     To summarize, we have used a consistent quantum theoretical description of the magnetic spin-flip scatterings
     of a neutral two-level atom trapped in the vicinity of a superconducting body. We have derived a simple scaling law for 
     the corresponding spin-flip lifetime for a superconducting thick slab.
     For temperatures below the superconducting transition temperature $T_c$, the lifetime
     has been found to be enhanced by several orders of magnitude in comparison to the case of a normal conducting slab.
     We believe that this result represents an important step towards
     the design of atomic chips for high-quality quantum information processing.

     We are grateful to Heinz Krenn for helpful discussions. 
     Work supported in part by the Austrian Science Fund (FWF).

%%%%%%%%%%%%%%%%%%%%%%%%%%%%%%%%%%%%%%%%%%%%%%%%%%%%%%%%%%%%%%%%%%%%%%


\begin{thebibliography}{xx}
\bibitem{remark_local}
    Strictly speaking, a local dielectric response is only valid if, for a given temperature $T$, the skin depth $\delta(T)$ 
    associated with the normal conducting part of the current density is sufficiently large in comparison to the mean free 
    path $\ell(T)$ of the electrons and  %i.e. $\delta(T) \gg l(T)$, and 
    the penetration depth $\lambda(T)$ of the field large in comparison to the superconductor coherence length $\xi(T)$. %, i.e. $\lambda(T) \gg \xi(T)$. 
    Superconductors satisfying the latter condition are known as London superconductors.
\end{thebibliography}
\end{document}